\documentstyle[aps,prb,epsf]{revtex} 

\begin{document} 

%\draft

\title{Scaling behaviour of the transmission poles for Dirac comb
}

\author{A. Korm\'anyos, J. Cserti, G. Vattay\\ 
E{\"o}tv{\"o}s University, Department of Physics of 
Complex Systems, \\
H-1117 Budapest, P\'azm\'any P{\'e}ter s{\'e}t\'any 1/A, Hungary}

%\wideabs{

\maketitle

\begin{abstract}
The transmission poles of $N$ number of identical Dirac delta potentials placed
periodically in one-dimension are examined in the complex-energy plane. 
The numerical results show that the imaginary part
of the poles scales with $1/N$. An approximate form of the poles is
derived which supports the scaling behaviour of the poles found numerically.
It is shown that the imaginary part of the poles are proportional to
their real part for the poles close to the ends of the bands.  
\end{abstract}

%}

\section{Introduction}\label{bev}

In recent years, quantum transport in mesoscopic systems has become a rapidly
developing research area\cite{Carlo-Houten,Fukuyama,Imry-konyv,Datta}.
In the pioneering papers by Landauer and B\"uttiker it was
shown that the conductance of such devices can be related to the transmission
coefficients depending on the scattering properties of the 
system\cite{Landauer,Buttiker}.
In electron waveguide structures the resonant behaviour and the transmission
phenomena have become an extensively studied 
problem\cite{polus-Shao,reso}. 
The transmission resonance caused by the quasi-bound states 
are in close connection with the poles of the transmission amplitude in the
complex-energy plane. The quasi-bound state at energy $E_0$ and with decay time
$\tau=\hbar /\Gamma$ corresponds to a simple pole in the transmission
amplitude $t(z)$ at the complex-energy $z=E_0 -i \Gamma /2$ and the energy
dependence of the transmission
probability $T(E)=|t(E)|^2$ for $\Gamma \ll E_0$ is given by 
the well-known Breit-Wigner formula\cite{Breit-Wigner-1,Breit-Wigner-2}
\begin{equation}
T(E) = \frac{\Gamma^2 /4}{{\left( E-E_0 \right)}^2 + \Gamma^2 /4}.
\end{equation}

The pole structure of the transmission of a one-dimensional periodic array 
of Dirac delta potential (Dirac comb) is investigated in this paper.
The transmission amplitude of N unit cells is expressed by the
transmission of one unit cell. The poles of the transmission amplitude 
are calculated numerically and a definite structure of the
poles in the complex energy plane shows up.
An analytical expression is derived for the imaginary part of 
the poles and it is shown that it scales with $N^{-1}$, where $N$ 
is the number of unit cells in the system. The agreement with the numerical
results are excellent.    

The rest of the paper is organised as follows. In section 
\ref{transfer} using the transfer matrix method the transmission
coefficient of $N$ number of unit cells is derived for the case of Dirac comb.
In section \ref{num-eredmeny} the numerical calculation of the poles are 
presented. In section \ref{anal-eredmeny} an approximate expression of the
poles are derived and compared with the numerical results.
Finally, in section \ref{veg} our conclusions are given.

\section{The transfer matrix approach}\label{transfer}

In this section the transmission of an one-dimensional periodic array of 
$N$ Dirac delta potentials is considered. 
In one-dimension the single-channel description of the scattering
process is valid.  In this case Sprung $et.\ al.\ $\cite{Sprung} 
derived the following compact expression for the transfer matrix of 
$N$ number of periodically placed identical cells in terms of the 2x2 transfer 
matrix ${\bf M}$ of a single-cell:
\begin{equation}
{\bf M}^{N}=\frac{1}{\sin{{\Phi}}}\,
[{\em {\bf M}}\sin{N{\Phi}}-{\bf 1}\sin{(N-1){\Phi}}],
\label{Tnedikenen}  
\end{equation}
where
$\cos{\Phi}=1/2 \,{\rm Tr}\,{\bf M}$.
Similarly, the following simple expression holds for\cite{Sprung} 
the transmission probability $T_N=|t_N|^2$ of the $N$-cell array: 
\begin{equation}
\frac{1}{|{t_{N}}|^{2}}=1
+\frac{\sin^{2}N{\Phi}}{\sin^{2}{\Phi}}\left(\frac{1}{|{t_{1}}|^{2}}-1\right),
\label{T_N}
\end{equation}
where $t_1$ is the transmission amplitude of a single-cell and
the Bloch phase $\Phi$ associated with the
infinite periodic potential is given by 
\begin{equation} 
\cos{\Phi}={\rm Re}\,(1/t_1),
\label{cosfi}
\end{equation} 
Note that the transmission amplitude depends on the energy of the electron
involved in the scattering process.

These results can be applied for a one-dimensional periodic array of $N$
Dirac-delta potentials. The transmission amplitude $t_1$ for the case of 
unit cell containing a Dirac delta potential of strength $\lambda$ and 
located in the centre of the unit cell is given by
\begin{equation}
t_{1}=\frac{k}{k+i{\lambda}/2}e^{ikd},
\label{t1}
\end{equation}
where $k=\sqrt{E}$, the length of the unit cell is denoted by $d$
and $E$ is the energy of the electron.  Hereafter we use the $\hbar = 2
m = 1$ units. 
Using Eq.\ (\ref{cosfi}) we find 
\begin{equation}
\cos \Phi(E)=\cos kd
+\frac{\lambda}{2k}\sin kd .
\label{cosfi-Dirac}
\end{equation}

The quasi-bound states of the system are given by the poles of 
the transmission amplitude in the complex energy
plane\cite{polus-Shao}.
Hence, for the $N$-cell system $1/T_N =0$ which, 
using Eq.\ (\ref{T_N}), can be written as 
\begin{equation}
1+\frac{\sin^{2}N{\Phi}}{\sin^{2}{\Phi}}
\left(\frac{1}{|{T_{1}}|^{2}}-1\right)=0.
\end{equation}
Thus, substituting  Eq.\ (\ref{t1}) into the above equation 
one finds that for Dirac comb the complex energy $E$ 
of the quasi-bound states is the solution of the following  equation:
\begin{equation}
-\frac{4E}{{\lambda}^{2}}=\frac{\sin^{2}N{\Phi(E)}}{\sin^{2}{\Phi(E)}}
\label{kotott-allapot}
\end{equation} 
where $\Phi(E)$ is given by Eq.\ (\ref{cosfi-Dirac}).

\section{Numerical results}\label{num-eredmeny}

\begin{figure}
{\centerline{\leavevmode \epsfxsize=8.5cm \epsffile{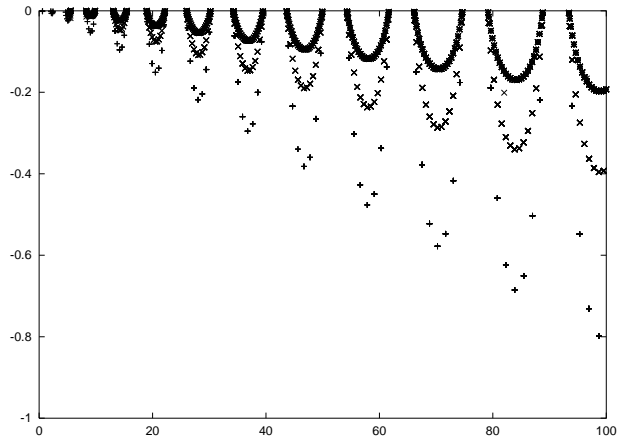 }}}
\caption{The poles of the transmission amplitude of the Dirac comb for
$N=8, 16, 32$ number of Dirac deltas (the poles are denoted 
by plus signs, crosses and stars, 
respectively). The strength of the Dirac delta 
potentials, $\lambda=10.0$, and the distance between them, $d=4$.
\label{abra-polusok}}
\end{figure}

We now present the numerical solution of  Eq.\ (\ref{kotott-allapot}).
In Fig.\ \ref{abra-polusok} 
the poles in the complex energy plane are shown for different number of
Dirac delta potentials. 
It can be seen that increasing the number of Dirac delta potentials, $N$ 
in the array the poles move towards the real energy axis.
It is also clear from the figure that the real part of the poles are 
separated into allowed and forbidden bands. The formation of the
band structure of the infinite periodic array (Kronig-Penney 
model\cite{Ashcroft}) can be
seen even for $N=8$.  In each allowed band there are $N-1$ number of
poles. 
The width of the forbidden bands decreases with increasing the real part 
of the complex energy.   
The real part of the poles, $E^m_l$ can be given by 
\begin{equation}
\Phi (E^m_l) = l \pi /N, 
\label{fi0}
\end{equation}
where $l=1,\dots N-1$ and $\Phi(E)$ is given by Eq.\ (\ref{cosfi-Dirac}). 
Hence, the real part of the $l$th poles in the $m$th band, $E^m_l$ 
is determined by 
\begin{equation}
\cos k_{lm}d+\frac{\lambda}{2k_{lm}}\sin k_{lm}d =\cos (l\pi/N),
\label{E0}
\end{equation}
where $k_{lm}=\sqrt{E^m_l}$.

\section{Scaling of the poles}\label{anal-eredmeny}

In order to see the scaling behaviour of the poles 
an approximate solution of 
Eq.\ (\ref{kotott-allapot}) for the quasi-bound states is derived.  

The solution of Eq.\ (\ref{kotott-allapot}) 
is a complex energy in the form of
$E=E^m_l -i \Gamma^m_l /2$, where $E^m_l$ is the 
$N-1$ number of distinct solutions of Eq.\ (\ref{E0}) indexed
by $l$ in the $m$th allowed band and the decay time of the quasi-bound
state is $\hbar /\Gamma^m_l$.  
Assuming $\Gamma^m_l /2 \ll E^m_l$, the Bloch phase $\Phi(E)$ defined by 
Eq.\ (\ref{cosfi-Dirac}), can be expanded
in first order in $\Gamma^m_l$ around  $E^m_l$:
\begin{equation}
{\Phi}(E) \approx {\Phi}(E_0)+\delta \Phi (E^m_l,\Gamma^m_l),
\end{equation}
where 
\begin{equation}
\delta \Phi (E^m_l,\Gamma^m_l)=
-i\left.\frac{d{\Phi(E)}}{dE} \right|_{E^m_l}\frac{\Gamma^m_l}{2}. 
\end{equation}

Finally, Eq.\ (\ref{kotott-allapot}) becomes
\begin{equation}
-\frac{4(E^m_l+i\Gamma^m_l/2)}{{\lambda}^{2}}=
\frac{\sin^{2}N({\Phi}(E^m_l)+i{\delta}{\Phi}(E^m_l,{\Gamma^m_l}))}
{\sin^{2}({\Phi}(E^m_l)+i{\delta}{\Phi}(E^m_l,{\Gamma^m_l}))} .
\label{polegy}
\end{equation} 
Substituting Eq.\ (\ref{fi0}) into Eq.\ (\ref{polegy}) and neglecting
$\delta \Phi $ in the denominator of the right hand side of 
Eq.\ (\ref{polegy}) results in an implicit equation for $\Gamma^m_l$ 
of the quasi-bound state:
\begin{equation}  
\frac{\Gamma^m_l}{2}=\frac{1}{N} {\rm arsinh} 
\left(\frac{2 \sqrt{E^m_l+i\frac{\Gamma^m_l}{2}} 
\sin\left(l\pi/N\right)}{\lambda}\right) 
{\left(\left.\frac{d{\Phi(E)}}{d{E}}\right|_{E^m_l}\right)}^{-1},
\label{gamma1}
\end{equation}
where the last factor can be determined from Eq.\ (\ref{cosfi-Dirac}):
\begin{equation} 
\frac{d{\Phi}(E)}{dE}=\frac{d}{4k\sin{\Phi}}
\left[\left(2+\frac{{\lambda}}{Ed}\right)\sin{kd}-
\frac{{\lambda}\cos{kd}}{k} \right],
\label{implderiv}
\end{equation}
and $k=\sqrt{E}$. For $\Gamma^m_l \ll E^m_l$ the implicit 
Eq.\ (\ref{gamma1}) can be solved 
iteratively for $\Gamma^m_l$. 
Substituting $\Gamma^m_l =0 $ into the right hand
side of Eq.\ (\ref{gamma1})  yields
\begin{equation}
\frac{{\Gamma}^m_l}{2}=\frac{1}{N}\,
\frac{4k_{lm}\sin \left(l\pi/N \right)}
{\left(\lambda /k^2_{lm}+2d\right) \sin k_{lm}d
-\left(\lambda d/k_{lm}\right)\cos k_{lm}d} \,
{\rm arsinh}\left(\frac{2 k_{lm} 
\sin\left(l\pi/N \right)}{\lambda}\right).
\label{gamma2}
\end{equation} 
The approximate solution of Eq.\ (\ref{kotott-allapot}) as the complex
energy, $E^m_l-i\Gamma^m_l/2$ of the $l$th quasi-bound state in the $m$th
band is given by Eqs.\ (\ref{E0}) and (\ref{gamma2}). 

\begin{figure}
{\centerline{\leavevmode \epsfxsize=8.5cm \epsffile{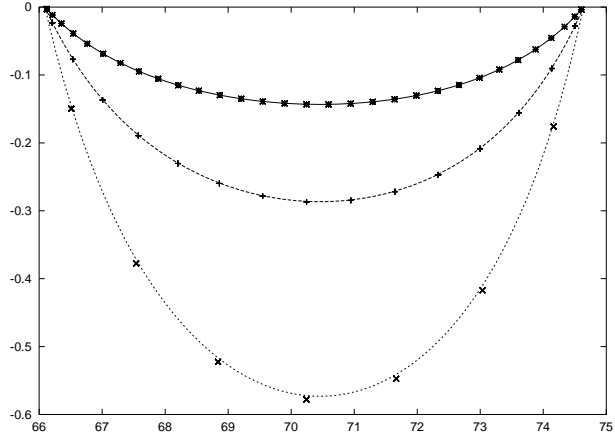 }}}
\caption{One band of the poles (an enlarged portion of
Fig.\ \ref{abra-polusok} for $66 \leq \rm{Real}\{E\} \leq 75$) and the 
plot of Eq.\ (\ref{gamma2}) are shown. 
The number of Dirac deltas are $N=8, 16, 32$ 
(the poles are denoted by plus signs, crosses and stars, 
respectively). 
\label{abra-polus-skala}}
\end{figure}

In Fig.\ \ref{abra-polus-skala} the numerically determined poles are shown
for  $N=8, 16, 32$ number of Dirac delta potentials (this is an 
enlarged portion of Fig.\ \ref{abra-polusok}) in the range of 
$66 \leq \rm{Real}\{E\} \leq 75$ together with the approximate analytical 
solution given by  Eqs.\ (\ref{E0}) and (\ref{gamma2}) when $E^m_l$ is
taken as a continuous variable.   
One can see that the envelope of the poles is well approximated by 
Eq.\ (\ref{gamma2}). 
The agreement between the analytic and the numerical
solutions is excellent. Similar agreements are found in the other bands.
The approximate solution becomes even better for increasing $N$.
A small deviation appears only in the middle of the band but increasing 
$N$ it is decreasing.
From Eq.\ (\ref{gamma2}) one can see that the imaginary part of the 
complex energy of the poles are proportional to $1/N$.  
It is easy to show that the factor multiplied by $1/N$ 
in Eq.\ (\ref{gamma2}) does not depend on $N$ in the leading order of 
$1/N$. Therefore, for large $N$ 
the imaginary part of the poles scales as  
\begin{equation}
\Gamma^m_l \sim 1/N
\label{pol-scale}
\end{equation}   
for all the poles.

\begin{figure}
{\centerline{\leavevmode \epsfxsize=8.5cm \epsffile{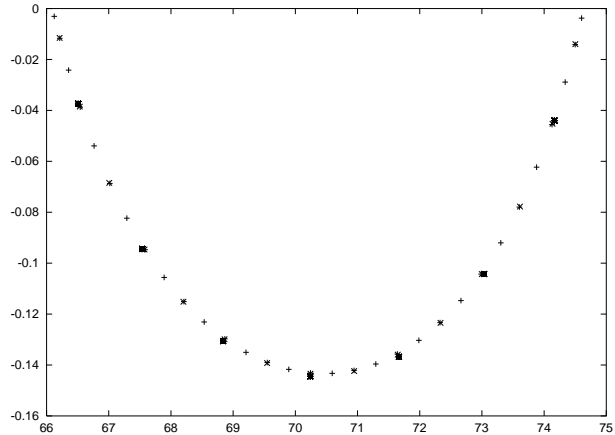 }}}
\caption{$N\Gamma^m_l/2$ as a function of \rm{Real}\{E\} for the case of 
$N=8, 16, 32$ number of Dirac deltas.
The same parameters of the system and signs are used as 
in Fig.\ \ref{abra-polusok}.
\label{abra-polus-egybe}}
\end{figure}

Using the numerical solutions of  Eq.\ (\ref{kotott-allapot}) 
in  Fig.\ \ref{abra-polus-egybe} $N\Gamma^m_l/2$ is plotted as a 
function of the real part of the poles for $N=8, 16, 64$. 
As it is seen from the figure the numerically found poles follows the
scaling behaviour of the poles given by  Eq.\ (\ref{pol-scale}).   
Again a very small deviation arises only at the middle of the bands for $N
\le 16 $.

\begin{figure}
{\centerline{\leavevmode \epsfxsize=8.5cm \epsffile{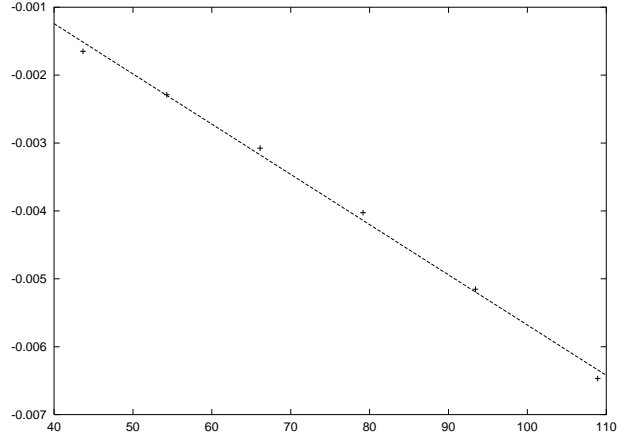 }}}
\caption{$\Gamma^m_l/2$ as a function of $E^m_l$  for $N=32$ and $l=1$, 
and the plot of Eq.\ (\ref{pole-energia}).
The same parameters of the system are used as 
in Fig.\ \ref{abra-polusok}.
\label{abra-polus-energia}}
\end{figure}

Using $\rm arsinh(x)\approx x$ for $x \ll 1$ 
(in this case $\sin \Phi(E^m_l) \ll 1$)
in Eq.\ (\ref{gamma2}) one can show that 
\begin{equation}
\frac{\Gamma^m_l}{2} =  E^m_l\frac{8}{\lambda N}\,
\frac{\sin^2 \left(l\pi/N \right)}
{\left(\lambda /k^2_{lm}+2d\right) \sin k_{lm}d
-\left(\lambda d/k_{lm}\right)\cos k_{lm}d}. 
\label{pole-energia}
\end{equation}
for fixed $N$ and $l \ll N$ or $N-l \ll N$ (these poles 
are close to the end of the band $m$). 
Again it can be shown that the factor multiplied by $E^m_l$ 
in Eq.\ (\ref{pole-energia}) is weakly dependent on the band index
$m$ for those values of $E^m_l$  when all the terms in 
the denominator of the right hand side of Eq.\ (\ref{pole-energia}) is
negligible except for the term $2d\sin k_{lm}d$. 
Thus, in the complex energy plane 
the poles at the end of each band
move down from the real axis as the band index $m$ increases. This
tendency can also be seen clearly from Fig.\ \ref{abra-polusok}. 
However, from  Eq.\ (\ref{pole-energia}) a more rigorous statement can
be made, namely at fixed $N$ and 
poles with $l$ close to the end of the band $\Gamma^m_l \sim E^m_l$.
In Fig.\ \ref{abra-polus-energia} the imaginary part of the poles,
$\Gamma^m_l$ calculated numerically is plotted as a
function of the real part of the poles, $E^m_l$ for $l=1$ as well as
the analytic form given by Eq.\ (\ref{pole-energia}).   
The numerically calculated poles are in good agreement with the
analytic form given by Eq.\ (\ref{pole-energia}) even for $N=8$. 
The poles move down
from the real axis in a way that their imaginary part are proportional
to their real part for those poles which are close to the ends of the
bands.

\section{Conclusion}\label{veg}

We have studied transmission properties of the one-dimensional 
periodic array of Dirac delta potential. Using the transfer matrix
method a closed form of the transmission amplitude is given for
finite number of Dirac delta potentials (finite number of unit cells
in the Dirac comb). 
The quasi-bound states, which affect on the transmission
properties of the system, are related to the poles of the transmission
amplitude.
The pole structure of the system  is investigated for different number
of unit cells in the Dirac comb. 

The numerical solutions reveal a definite structure of
the poles. The poles are well separated into allowed and forbidden 
bands much in the same way as in the Kronig-Penney model but the 
'pole-spectrum' is not continuous. Each band contains $N-1$ number of
poles, where $N$ is the number of unit cells in the system.  
To check our numerical results an analytical form for the 
imaginary part of the poles has been derived. Although this is an 
approximate expression the comparison with the numerical results
shows an excellent agreement especially for large $N$.
From our analytic solution it has also been proved that the 
imaginary part of the poles is proportional to $1/N$. 
Furthermore, it has been demonstrated that for the poles close 
to the ends of the bands the imaginary part of 
the poles are again proportional to the real part of the poles.

\acknowledgements
This work was supported by the EU.\ TMR within the programme 
``Dynamics of Nanostructures'' jointly with OMFB, the Hungarian 
Science Foundation OTKA  T025866, 
the Hungarian Ministry of Education (FKFP 0159/1997).

%\bibliographystyle{prsty}
%\bibliography{/localnet/robin/cserti/tex/cikkek}
%\bibliography{/home/cserti/tex/cikkek}

\end{document}